\documentclass[preprint,12pt,review,authoryear]{elsarticle}

\usepackage{amssymb}
\journal{Planetary and Space Science}

\begin{document}

\begin{frontmatter}



\title{Meteoroid structure and fragmentation}


\author{M. D. Campbell-Brown}
\ead{margaret.campbell@uwo.ca}
\address{University of Western Ontario, Department of Physics and Astronomy, London, ON, Canada N6A 3K7 +15196612111 }

\begin{abstract}

The physical composition and structure of meteoroids gives us insight into the formation processes of their parent asteroids and comets. The strength of and fundamental grain sizes in meteoroids tell us about the environment in which small solar system bodies formed, and the processes which built up these basic planetary building blocks. 
The structure of meteorites can be studied directly, but the set of objects which survive entry through the atmosphere is biased toward large, strong objects with slow encounter speeds. Fragile objects, small objects and objects with high relative speeds are very unlikely to survive impact with the atmosphere. Objects between 100 microns and 1 meter, which are not strong enough to survive the ablation process, must be studied by radar or optical methods. 

Large meteoroids, which produce bright fireballs, are generally studied by investigating their compressive strength when they fragment, and their strength can also be inferred indirectly from their end heights. 

Fragmentation in faint meteors can be inferred from interference in radar observations, or observed directly with high-resolution optical systems. Meteor light curves, begin heights and time-evolving spectra can also be used to infer meteoroid structure. 

This paper presents a review of the meteor observation methods currently used to infer the structure and fragmentation of meteoroids in the millimeter to meter size range, and the current state of understanding these observations have given us.

\end{abstract}


\end{frontmatter}

%\linenumbers

\section{Introduction}
\label{intro}

The structure of meteoroids is scientifically interesting for a variety of reasons. Meteoroids represent, in most cases, unprocessed material from the formation of the solar system, and can provide insights into the formation of their parent asteroids and comets. This in turn can help provide constraints for models of solar system formation, and help us understand conditions in the early solar system when the planets were forming. In some cases, the effects of exposure after separation from a parent body may become important as well, particularly for meteoroids which approach the Sun. 

Small meteoroids, which do not reach the ground, may nevertheless pose a hazard to spacecraft, and their strength and structure determine the type of damage they will do in the event of a collision. 

The most obvious way to study the structure of unprocessed solar system material is to look at meteorites, which have survived passage through the atmosphere. This allows limits of strength and density to be determined, from the very strong material represented by the largest surviving fragment of the Chelyabinsk meteorite \citep{borovicka2013} and the Carancas impactor \citep{borovicka2008} to the very weak carbonaceous chondrites like Tagish Lake \citep{borovicka2015b}. Studying meteorites in the lab also allows fundamental grain sizes and compositions to be measured. Meteorite studies do provide an important basis for understanding material in the early solar nebula, but they suffer from very strong observing biases: only large, slow, strong material will survive entry through the atmosphere, and the strongest material is much more likely to be collected. Objects smaller than 10 cm (including most cometary material), or with speeds higher than about 20 km/s (again, including most cometary material) will be completely ablated in the atmosphere. 

The best way to study this cometary material is through dedicated spacecraft visits to comets. Material from one comet (81P Wild 2) has been returned to Earth as part of the Stardust mission. The grains were collected at a relative velocity of just over 6 km/s, so only the largest particles generally yielded thermally unprocessed material, but the overall strength of the particles could be evaluated based on the shape of the tracks. Strong particles produced long, narrow cavities in the aerogel, while fragile clusters produced wide, short cavities. The most unexpected finding from Stardust was the presence of strongly heated material, not thought to be present in the cold region of the solar system where the comet formed \citep[See][ for a review]{brownlee2014}.

The Stardust data are of great importance, but they sample only grains up to 100 microns due to the limited collecting area and exposure time of the detector, and most of the mass lost by the comet was in grains larger than this. The Rosetta mission's COSIMA mass spectrometer and microscope (COSISCOPE) was able to photograph more than 10,000 grains with 10 $\mu$m resolution \citep{langevin2016} on nine 1 cm$^2$ targets over the course of a year. The dust particles struck the targets at very low speed, and while there was deformation of fragile particles, there was no significant thermal heating. The largest particles were several hundreds of microns in size, and some were consistent with being smaller components of millimeter sized objects which fragmented before impact. At large solar distances, material from the comet was mainly fragile clusters, while closer to the sun nearly a third of the grains were compact. There seemed to be fewer fragile clusters from the southern hemisphere, implying more sintering on the comet's southern surface \citep{merouane2017}. The dust production was higher before perihelion, unlike the gas production, which the authors attribute to the loss of volatile-poor dust, followed by volatile-rich dust as the surface layers were shed. The largest particles detected by COSIMA were close to the small end of conventionally detectable meteors. Preliminary analysis with the ToF-SIMS mass spectrometer on COSIMA has produced evidence of a small calcium-aluminum inclusion, similar to the high-temperature inclusions seen in the Stardust material \citep{paquette2016}. 

The MIDAS Micro-Imaging Dust Analysis System on Rosetta used an atomic force microscope to image micron-sized comet dust, and found that even these very small grains were hierarchical clusters of smaller grains, suggesting that even the component grains observed by COSIMA may themselves be clusters of smaller grains \citep{bentley2016}. This fresh dust from the comet may change with long exposure to radiation off the comet's surface, but is a good starting point for understanding cometary meteoroids.

Dedicated spacecraft missions can provide important information, but they are expensive, and the best way to study material from a wide range of comets and asteroids is to observe material as it ablates in the atmosphere. We review below the current techniques and results.

\section{Fireball observations}

Bright fireballs include the largest shower meteors, mainly from cometary sources, and also asteroidal material. They are typically observed with very wide field or all-sky cameras, since the flux of these large objects is low and requires a large collecting area. 

\subsection{Direct fragmentation measurements}

While wide field cameras typically used for observations of bright meteors have low spatial resolution, bright meteors often show spread between fragments of several kilometers \citep{brown1994, borovicka2003}, making fragmentation easy to characterize. 

An important way to probe the structure of large meteoroids is to examine the dynamic pressure needed to cause fragmentation. This only applies to large meteoroids which penetrate deep in the atmosphere: faint meteors ablate at such high altitudes that pressure is negligible compared to even the weakest contact forces. \citet{popova2011} review the bulk strength of observed meteorite falls and the Prairie, European and Satellite fireball networks, along with the Carancas impact. The majority of the strengths calculated fall between 0.02 MPa (for first fragmentation events) and 10 MPa. In general, measured strengths of fireballs are significantly less than the strength of recovered meteorites, which is typically 30 MPa or greater \citep[e.g.][]{cotto2016}. They conclude that pre-entry, meter-scale meteoroids are fractured or rubble piles. Only the Carancas impact and the final fragment of Chelyabinsk \citep{borovicka2013} had strengths consistent with monolithic chondrites. 

\citet{borovicka2015a} searched particularly for large meteoroids bound only by weak contact forces, as expected for a rubble pile. He found no evidence in Almahata Sitta, Bene\v{s}ov or Chelyabinsk that fragmentation occurred at less than 100 Pa, and concluded that even large, heterogeneous meteoroids are bound by stronger forces. He notes that early fragmentation at pressures of about 25 Pa may not be easy to observe, so rubble-pile meteoroids cannot be ruled out in general.

\subsection{Indirect evidence for structure: End height criteria}

For bright meteors, the height at which they stop producing light is a function of their speed, size, and entry angle, and also their structure and composition. \citet{ceplecha1976} attempted to control for the first three factors with the $PE$ criterion, which sorts meteors into groups based on:
\begin{equation}
PE=\log \rho_E +A\log m_\infty+B \log V_\infty+C\log(\cos z_R)
\end{equation}

\noindent where $\rho_E$ is the atmospheric density at the end height, $m_\infty$ is the total mass (calculated photometrically), $V_\infty$ the out-of-atmosphere speed, and $z_R$ the angle between the radiant and zenith. They used this parameter to divide fireballs into four groups (I, II, IIIa and IIIb), which they associated with ordinary chondritic material, carbonaceous chondritic material, and strong and weak cometary material. The divisions were based on perceived groupings of fireballs on a histogram of $PE$ values for meteors observed with the Prairie Network.

The $PE$ criterion has the advantage that it can be calculated for any observed fireball. A more precise measure of the structure of meteoroids is the $SD$ parameter, also formulated in \citet{ceplecha1976}, which uses the geometry of the particle and the ablation coefficient:
\begin{equation}
SD=\log\left(\Gamma A\rho_m^{-2/3}\right)+\log\left(\frac{\Lambda}{2\Gamma\zeta}\right)
\end{equation}

\noindent Here, $\Gamma$ is the drag coefficient (fraction of momentum transferred from the airstream to the meteoroid), $A$ is the shape factor, $\rho_m$ is the meteor density, $\Lambda$ is the heat transfer coefficient (fraction of kinetic energy transferred from the airstream to the meteoroid), and $\zeta$ is the heat of ablation. These parameters require very precise observations, since they rely on accurate deceleration measurements. $SD$ and $PE$, when they can both be calculated, are strongly correlated and tend to sort fireballs into the same groups.

\citet{ceplecha1980} tried again for an end-height criterion which could be easily calculated from observed quantities, this time removing the assumptions which go in to calculating the photometric mass. The resulting criterion was the $AL$ (ablation-light) parameter:
\begin{equation}
AL=5\log v_\infty+2\log\left(\frac{\rho_E}{\cos z_R}\right)-0.83\log\left(\int_{t_B}^{t_E}Idt\right)
\end{equation}

\noindent where $t_B$ and $t_E$ are the begin and end times. The $AL$ parameter is also strongly correlated with the $PE$ criterion. It is worth noting that the integrated luminous intensity still depends on the luminous efficiency, which may be different from one meteor to another, but it makes no explicit assumptions about that parameter.

These criteria can help determine if a particular meteoroid is very strong/ refractory or weak/volatile, but care should be taken in assigning compositions based on any of them. The distinctions between categories are not sharply defined, and the associations between categories and compositions are somewhat arbitrary, particularly for cometary material for which no recovered samples survive from observed fireballs. Observational errors can also move a specific meteor from one category to another.

\begin{figure}
 \caption{Tisserand parameter (with respect to Jupiter) and $PE$ parameter for fireballs observed with the MORP network. \label{fig:MORP_Tj_Pe}}
\includegraphics[width=35pc]{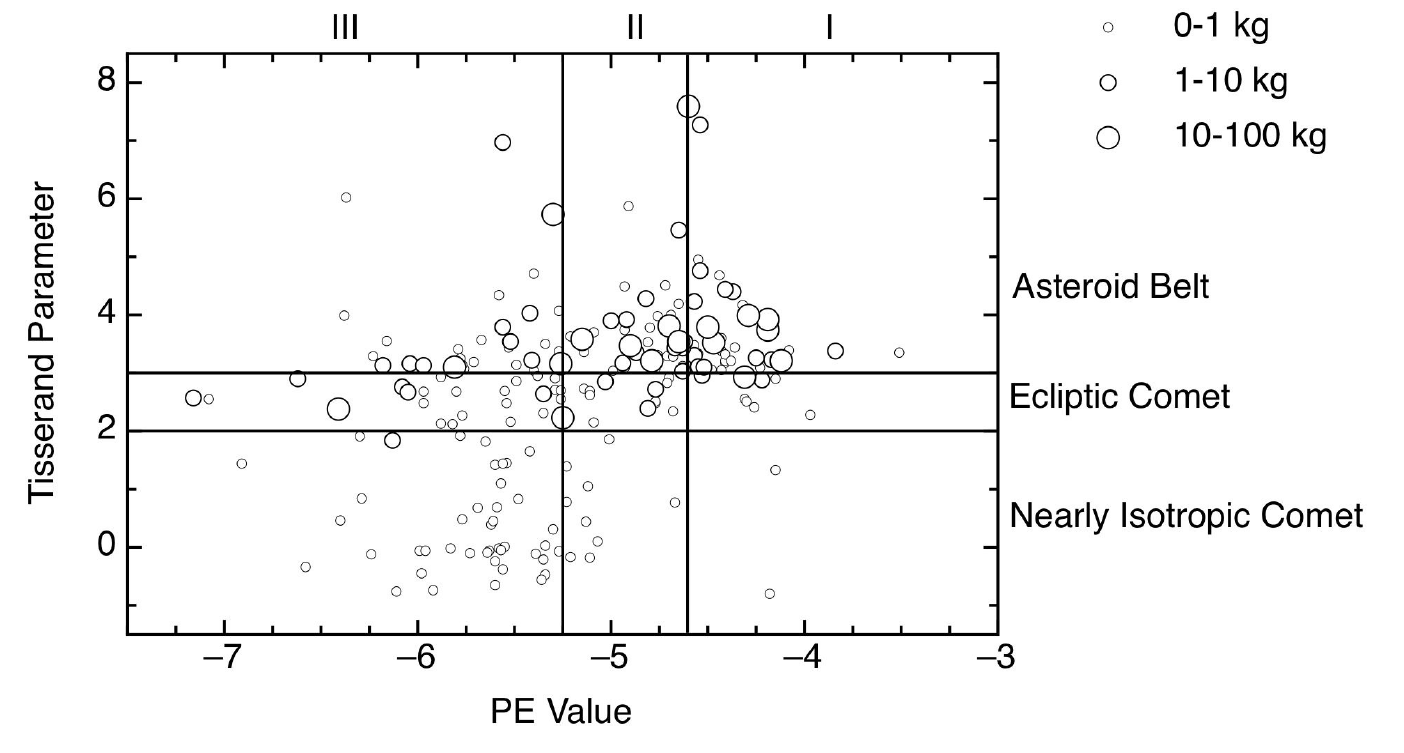}
\end{figure}

\citet{milley2010} examined the relationship between the $PE$ criterion and the Tisserand parameter with Jupiter for 638 meteors in the clear-sky survey of the Meteorite Orbit and Recovery Project (MORP) (Fig.~\ref{fig:MORP_Tj_Pe}). The Tisserand parameter is invariant under gravitational perturbations from Jupiter, and in the absence of significant radiation forces or close planetary encounters, it gives the population of origin (Asteroidal, Jupiter Family Comet or Nearly Isotropic Comet). The figure shows the expected concentration of strong (type I) meteoroids in asteroidal orbits and weak (type III) meteoroids in nearly isotropic orbits, but there are also a significant number of weak meteoroids in asteroidal orbits, and two strong meteoroids in clearly cometary orbits. Objects of this last type are of particular interest; \citet{borovicka2007} notes that strong objects (with compressive strengths of about 1 MPa) on long-period cometary orbits are rare, but present. The Karl\v{s}tejn fireball, for example, showed a lack of sodium and much larger strength than typical cometary material. This material may be fragments of the crust of comets, built up by irradiation while the comets were in the Oort cloud.

 \subsection{Modelling}

In cases where it is difficult to determine the number of fragments because they are too numerous, or observations are insufficient, the fragmentation behaviour of bright meteors may be determined by modelling the motion and light production of fireballs. The FM (Fragmentation Model) of \citet{revelle2002} models the deceleration and light curve of bright meteors. Any number of fragmentation events, producing either a cluster of small grains or a discrete number of large fragments, can be added at any point along the curve. The amount of mass lost by the meteoroid in each event can be estimated, which gives an idea of how much of the meteoroid disrupted: this varies from just a few percent to more than 90\% \citep{ceplecha2005}. 

The FM model has been very successful at matching the densities of fireballs with known associated meteorites \citep{ceplecha2005}, and typically finds that type I fireballs have the highest densities, consistent with ordinary chondrites, while type II are slightly less dense, and types IIIa and IIIb are the least dense, consistent with very porous material. 

Modelling can also point to deficiencies with other methods of categorizing meteors by their strength. The ablation coefficient $\sigma$, defined in terms of the heat transfer coefficient, heat of ablation and drag coefficient, is:

\begin{equation}
\sigma=\frac{\Lambda}{2\zeta\Gamma}
\end{equation}
The apparent ablation coefficient (not taking fragmentation into account) is sometimes used instead of the end height criteria to classify fireballs, since meteors which fragment heavily will have a much larger apparent ablation coefficient. In the case of the Tagish Lake fireball, however, the apparent ablation coefficient would indicate that the fireball was type II, while the fragmentation behaviour classified it as weaker than a type IIIb \citep{ceplecha2007}. The meteorites recovered from the fireball were indeed very low density and fragile.

\section{Faint meteor observations}

\subsection{Direct fragmentation measurements}

 It is possible to observe fragmentation directly in some cases: in radar observations, fragmentation can be inferred from the shape of a radar echo, and optical observations can image individual fragments or spreading clusters of debris. Direct measurements can be used to test indirect methods of inferring meteoroid structure, which are described in the following sections.

\subsubsection{Radar observations}

\citet{elford2004} used a technique called Fresnel holography to infer the presence of multiple fragments in a single meteor echo. He used the Buckland Park 54 MHz radar to observe trail echoes of meteors (echoes from the extended line of ionization produced by the meteor), and transformed the echoes from amplitude and phase versus time to electron density as a function of distance behind the meteor head. Some meteors showed several local maxima in the exponentially decaying trail of ionization behind the head, which was interpreted to be separate fragments trailing the main meteoroid. In some cases, separations as small as 10 cm along the line of sight could be measured. The secondary peaks in electron density are very small, which may mean they are difficult to distinguish from noise in many cases.

The presence of fragmentation has also been inferred from oscillations in head echo signal strength from high-power, large aperture radars, which scatter radiation from the dense ionization immediately surrounding the ablating meteoroid. \citet{campbell2007} modelled fragmenting meteors to show that oscillations in signal strength observed in ALTAIR radar meteors were consistent with meteoroids disrupted into many closely-spaced fragments. \citet{mathews2010} concluded that the majority of meteors observed by the Arecibo radar were fragmenting, based on fluctuations in the echo strength. The most convincing evidence for fragmentation in radar observations comes from the multistatic EISCAT high-power, large aperture radar \citep{kero2008}. In a radar with a single receiver station, fluctuations in head echo strength may be caused by the instrument, or by inherent variation in the signal strength of the meteoroid, caused, for example, by a fluctuating ablation rate. EISCAT has three receivers observing a common volume. The line-of-sight speed with which fragments separate will therefore be different for the three different stations. This allows echoes whose amplitudes fluctuate because of interference among multiple fragments to be separated from those which vary because of fluctuations in the ablation rate: the latter will vary at the same rate on all receivers, but the former will be different depending on the line of sight. 

\subsubsection{Optical observations}

\begin{figure}
 \caption{Examples of fragmenting meteors observed by the Canadian Automated Meteor Observatory (CAMO) \label{fig:NarrowSamples}}
\includegraphics[width=35pc]{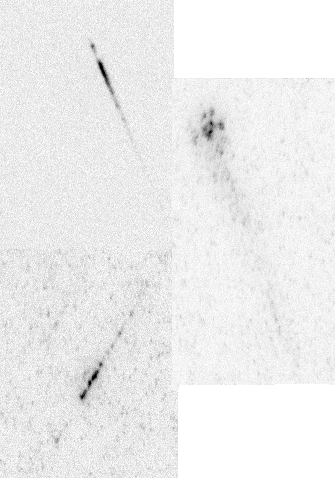}
\end{figure}

The resolution of optical systems can be determined by applying the angular pixel resolution of the system at typical meteor ranges observed by the system. It depends on the size of the image chip in pixels, the field of view of the system, and the distance to the viewing volume; it will vary for any given system between the closest and most distant parts of the observed volume. For all but all-sky systems, the resolution does not vary by more than a factor of two. Most intensified video systems have resolutions of 30 to 100 meters per pixel.  For example, the CILBO automated meteor observatory on the Canary Islands \citep{koschny2013} has a pixel scale of 2.3' per pixel (30 degrees and 768 pixels across), and an approximate range at the centre of the common observing volume (at 100 km) of 110 km, giving a spatial resolution of about 75 m per pixel. Meteors will appear as points in these systems unless they have fragment separations which are much greater than the resolution. \citet{fisher2000} found that only a small fraction of faint meteors show significant wake (physical spread of the light-producing area) in a standard system. The Canadian Automated Meteor Observatory (CAMO) \citep{weryk2013} has a tracking system which uses two cameras to track meteors. The wide field is used to detect the meteors, and a pair of mirrors then directs the meteors' light into a small telescope. The system has a spatial resolution as high as 3 meters per pixel, and generally better than 6 meters per pixel, depending on the range. The system tracks meteors smoothly and at a high frame rate (110 frames per second) so that meteor images have minimal smear. Examples of meteors observed with the CAMO tracking system narrow field are shown in Fig.~\ref{fig:NarrowSamples}.

Approximately 90\% of meteors observed with the CAMO system show obvious fragmentation \citep{subasinghe2016}, either in discrete fragments (about 5\% of meteors) or in the form of a long wake consisting of many fragments with different rates of deceleration. These wakes are likely due entirely to fragmentation: \citet{stokan2015} modelled the production of light around the head of meteors and found that the wake due to de-exciting atoms detached from the meteoroid was negligible.  The fragments are usually aligned with the motion of the meteor, but in a small number of cases the fragments show large transverse speeds \citep{stokan2014}, which may indicate explosive devolatilization of the meteoroids. 

CAMO meteors sometimes show a leading fragment, which decelerates less than the apparently fragmented material behind it. Such leading fragments have also been observed in wide-field systems, for example in a Leonid fireball, where the final fragment survived a dynamic pressure of about 2 MPa \citep{borovicka2000}. These events are evidence for large, strong constituent grains in cometary meteoroids, which has implications for mixing in the early solar system.

Even the 10\% of CAMO tracking meteors which do not show significant fragmentation may in fact be fragmenting: recent work \citep{campbell2017} has shown that even meteors with very short wakes cannot be fitted with a single body model, so a small amount of fragmentation may be taking place even in these short-wake cases.

 \subsection{Indirect evidence for structure}

Since direct measurements of fragmentation are more difficult for faint meteors than for fireballs, many ways of inferring structure from indirect evidence have been developed.

\subsubsection{Begin height and $k$ parameters}

Nearly all faint meteors fragment \citep[e.g.][]{subasinghe2016}, but the begin height gives some indication of the amount of energy needed to begin intensive ablation. \citet{ceplecha1967} came up with a parameter to characterize this value, which he called the $k_B$ parameter:
\begin{equation}
k_B=\log\rho_B+\frac{5}{2}\log v_\infty-\frac{1}{2}\log\cos z_R
\end{equation}
\noindent Here $\rho_B$ is the atmospheric density at the beginning of the observed trail (in g cm$^{-3}$), $v_\infty$ is the pre-deceleration speed (cm s$^{-1}$), and $z_R$ is the zenith angle. Ceplecha found that this parameter divided meteors observed with Super-Schmidt cameras into three broad groups, an A and a C group which showed up as prominent peaks in a histogram of $k_B$, and a plateau between them which he called group B. He then, using small-camera and intensified video datasets,  refined the classification as follows \citep{ceplecha1988}:

\begin{itemize}
\item	Asteroidal Meteors: 8.00$\le k_B$; ordinary chondrites
\item	Group A:  7.30$\le k_B<$8.00; carbonaceous chondrites (comets or asteroids)
\item	Group B: 7.10$\le k_B<$7.30; $q$ $\le$0.30 AU; Dense cometary material
\item	Group C1: 6.60$\le k_B<$7.10; $a <$5 AU; $i$ $\le$35$^\circ$; Regular cometary material
\item   Group C2: 6.60$\le k_B<$7.10; $a\ge$5 AU; Regular cometary material
\item	Group C3: 6.60$\le k_B<$7.10; $a<$5 AU;$i>35^\circ$; Regular cometary material
\item	Group D: $k_B<$6.60; Soft cometary material
\end{itemize} 

Offsets are needed to use this scheme on datasets other than Super-Schmidt: Ceplecha calculated that small-camera systems should subtract 0.30 from their calculated $k_B$ parameters to account for the less sensitive systems, while intensified video systems should add 0.15. \citet{kikwaya2011} found that it was necessary to add 0.18 to data from the CAMO influx cameras.

\begin{figure}
 \caption{Plot of begin height against speed for meteors observed with the CAMO influx system, with shower meteors marked. \label{fig:InfluxShowerHbV}}
\includegraphics[width=35pc]{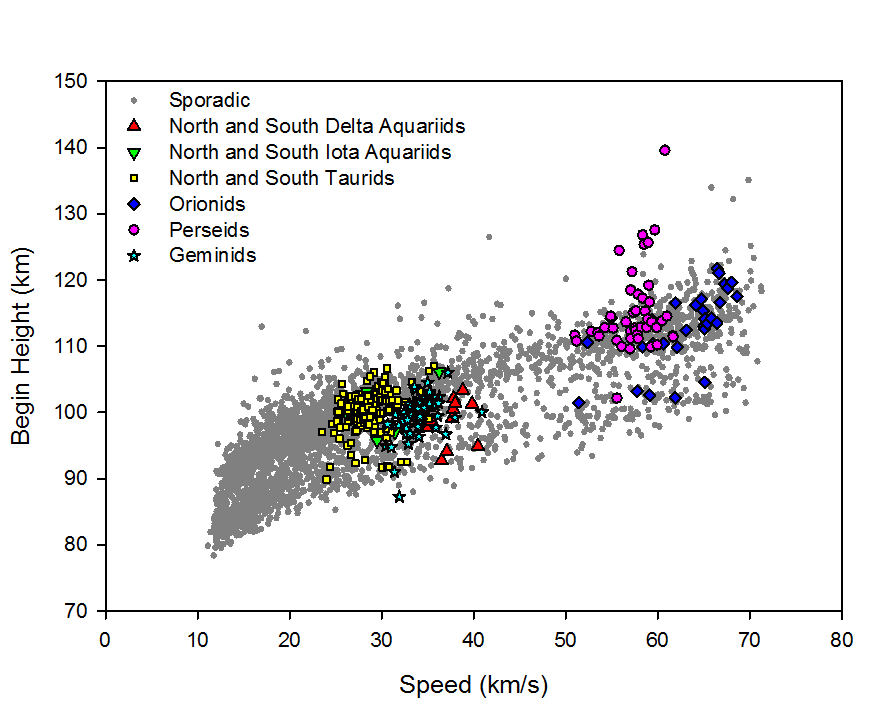}
\end{figure}

The effect of the $k_B$ parameter can be seen in a plot of begin height against speed, since the zenith angle has only a small effect. Fig.~\ref{fig:InfluxShowerHbV} shows data from the CAMO influx system, with shower meteors marked. The differences in strength or structure are obvious: sporadic meteors fall mainly along two lines, the top strip corresponding to Ceplecha's group C, and the bottom corresponding to group A. Most Perseids and Orionids are members of the top group (group C, with a few from group D), while most Geminids fall between the two lines, putting them in group B. A similar plot for CAMS data is given in \citet{jenniskens2016}, showing similar clustering of showers. For the Perseids, and for other fast showers, the begin height increases for high mass meteoroids because of sputtering, which artificially decreases the $k_B$ parameter. This can be seen in  Fig.~\ref{fig:InfluxShowerHbV} as a scattering of points above the top group.

\begin{figure}
 \caption{Plot of begin height against speed for meteors observed with the CAMO influx system, coloured by $k_B$ parameter. Note that the division between the top and bottom groups is less clearly given by the $k_B$ parameter at low speeds. \label{fig:HbV_kB}}
\includegraphics[width=35pc]{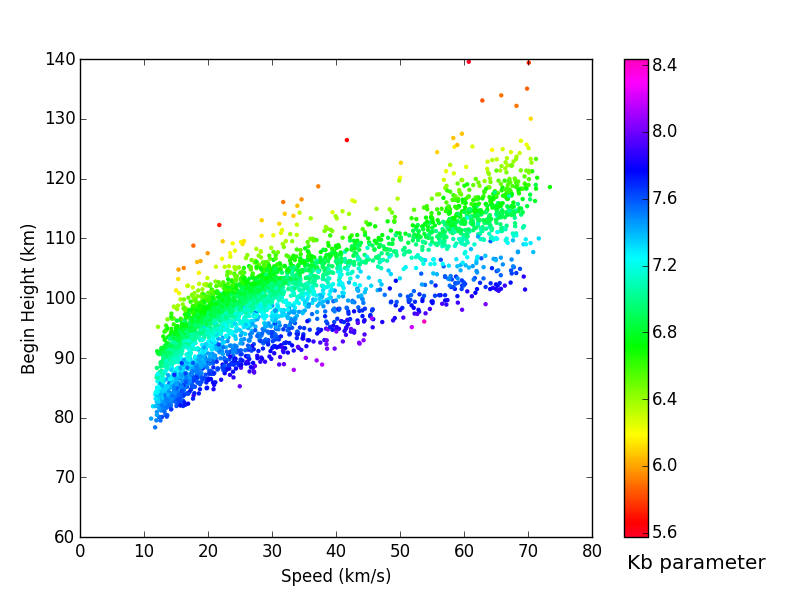}
\end{figure}

 Fig.~\ref{fig:HbV_kB} shows begin height against speed for meteors observed using the CAMO influx system, with points coloured according to their calculated $k_B$ parameter. At the lowest speed end, the $k_B$ parameter does not divide the two groups as cleanly as it does at high speeds.

\begin{figure}
 \caption{Plot of begin height against speed for meteors observed with the CAMS system, with lines of constant $k_c$ marked \citep{jenniskens2016}. \label{fig:CAMS_HbV_kc}}
\includegraphics[width=35pc]{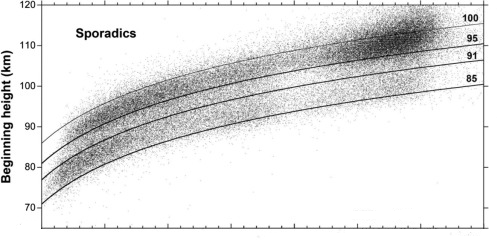}
\end{figure}

 \citet{jenniskens2016} points out that Ceplecha's $k_B$ parameter assumes that the temperature at the surface of the meteoroid depends on $v_\infty^{2.5}$, while the change in begin height with speed seems to follow $v_\infty^2$. They formulate a $k_c$ parameter:
\begin{equation}
k_c=H_b-(2.86-2.00\log v_\infty)/0.0612
\end{equation}
\noindent where $H_b$ is the begin height in km, and $v_\infty$ is the speed in km s$^{-1}$. Fig.~\ref{fig:CAMS_HbV_kc} shows a plot for sporadic meteors observed with the CAMS system from \citet{jenniskens2016}, marked with lines of constant $k_c$. These lines do a better job of dividing the different populations of begin heights.

 \subsubsection{Light curves and structure}

The light curves of meteors can be used to infer their behaviour during ablation. A homogeneous, non-fragmenting meteoroid will have a classical light curve with a peak toward the end of the curve. \citet{hawkes1975} predicted, in their dustball model in which small meteoroids disrupt into smaller fragments before the onset of luminous ablation, that these fragile objects should produce symmetric light curves, which had already been observed in Super-Schmidt data \citep{jacchia1955}. 

The F-parameter is often used to characterize the shape of light curves: it compares the position of the maximum to the begin and end of the observed meteor. It is calculated as the difference in between the height of maximum ablation and the begin height, divided by the difference between the begin and end heights. An F parameter less than 0.5 indicates that a meteor has an early peak; an F parameter of 0.5 indicates a symmetric light curve, and an F parameter between 0.5 and 1 indicates a late-peaked light curve. A classical, single body light curve has an F parameter of about 0.7.

\begin{figure}
 \caption{F parameter of Perseid meteors as a function of mass \citep{koten2004}. Large Perseids have light curves which peak late, while small Perseids have, on average, symmetric light curves. \label{fig:KotenFs}}
\includegraphics[width=35pc]{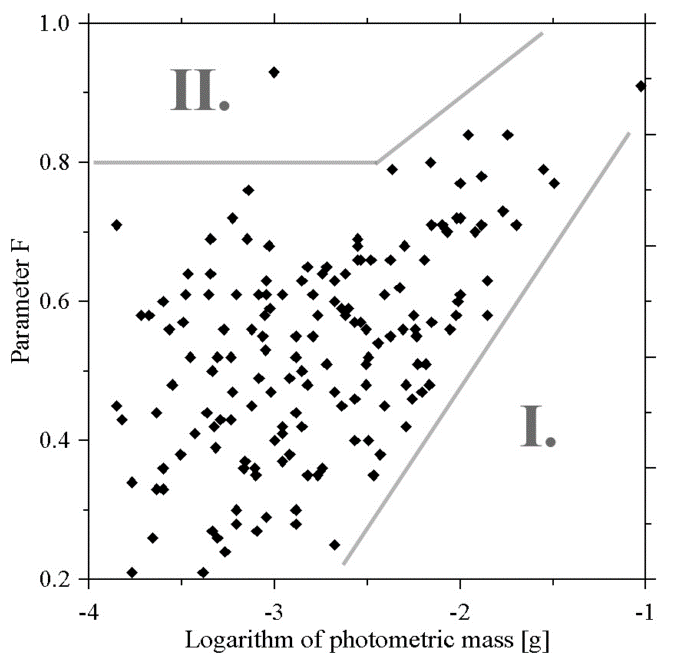}
\end{figure}

\citet{koten2004} and \citet{koten2015} measured F parameters for a number of meteor showers. The Draconids, which are commonly regarded as the most fragile of shower meteors, had the lowest average F parameter, though as with all showers the meteors spanned the full range of F values from very early peaked to very late peaked. The mode of the distribution was at F=0.35. The Leonids, also a fragile cometary shower,  had light curves which were nearly perfectly symmetrical on average, while other showers, including the Taurids, Perseids, Orionids, and Geminids had average F values between 0.53 and 0.59. All faint meteors have, on average, symmetric light curves, but larger meteoroids (Perseids larger than $10^{-5}$ kg, for example) have mostly late peaked light curves (see Fig~\ref{fig:KotenFs}). This is interpreted \citep[for example, by ][]{borovicka2006} as evidence that small meteoroids have completely disrupted before the onset of ablation, while larger meteoroids continue to fragment during ablation.

The link between fragmentation behaviour and light curves is more complicated than simple ablation theory predicts. In a study of high-resolution meteors observed with the CAMO tracking system, \citet{subasinghe2016} found that, while meteors which showed extensive fragmentation had, on average, symmetric light curves, so did meteors which showed very little evidence of fragmentation. In most cases, light curves are not a good predictor of fragmentation behaviour for faint meteors.

In particular, there is a class of slow, faint meteors with low begin heights which have very early-onset light curves \citep{campbell2015}. The high $k_B$ parameter of these particles implies that they are strong, but the sudden onset of luminosity implies disruptive fragmentation. Fourteen meteors with slow speeds, low begin heights and early-skewed light curves were observed in a spectral campaign by \citet{borovicka2005}; their spectra showed only iron lines. They were interpreted as being solid iron particles which melted and disrupted into droplets.

\subsubsection{Spectra}

Meteor spectra can be taken either by placing a diffraction grating in front of the imager \citep{borovicka2005}, or using notch filters to look at particular lines of interest \citep{bloxam2017}. The spectral lines present in meteoroids, combined with a model of the plasma, help to constrain the composition of meteoroids. The structure of meteoroids may be revealed in the way in which spectra change with time, indicating differential ablation of the minerals in the meteoroid, in other words ablation of different constituents at different points. The main lines visible in faint meteors are sodium, magnesium and iron. Some meteoroids show an early release of sodium compared to magnesium, and others show more uniform ablation, with all of the spectral lines showing the same brightness profile \citep{borovicka2006}. Differential ablation (mainly the early release of sodium) has been observed extensively in Leonids \citep{borovicka1999,murray1998}, but most Taurid meteors show uniform ablation \citep{borovicka2001}. \citet{borovicka2006} points out that differential ablation is not present in all meteors, and supposes that it may be linked to the fragmentation behaviour of meteoroids, in that disrupted meteoroids may completely fractionate and allow sodium to evaporate before other components of the meteoroid begin to ablate. However, \citet{bloxam2017} has found no link between fragmentation behaviour and differential ablation in a study of CAMO tracked meteors simultaneously observed with narrow-band spectral filters.

Some, but not all, meteors which approach the Sun closely (to within $\approx$0.2 AU) show a lack of sodium, which may indicate thermal processing; Geminid meteors in particular show low sodium emissions in many cases \citep{borovicka2005}. Recent modelling \citep{vojacek2018} suggests that sodium depletion is associated with both low perihelion and small grain sizes, while meteors with low perihelia requiring larger grains to model showed higher levels of sodium.

\subsection{Modelling and densities}

In principle, the density of a meteoroid can be determined by comparing the light curve and deceleration of the meteor to the equations of meteor ablation. In practice, the density determined this way depends on whether or not the meteor is assumed to fragment, and in what way it does so. \citet{babazhanov2002} and \citet{bellotrubio2002} both modelled the same set of Super-Schmidt photographic meteors; Babazhanov assumed the meteoroids underwent quasi-continuous fragmentation, while Bellot-Rubio et al. were able to fit nearly three quarters of the meteors with a single body model. Comparing meteors from different showers, both found that the Geminids had the highest density and the Perseids a much lower density, but Babazhanov's densities were generally higher than those found with the single body model, consistent with the increase in assumed surface area from fragmentation. 

Nearly all faint meteors fragment \citep{subasinghe2016}, so fragmentation cannot be discounted when determining meteoroid densities. The mechanism chosen for fragmentation (disruption or quasi-continuous fragmentation, for example) makes a difference to the density of best fit. 

\citet{kikwaya2011} did a thorough search of parameter space to find densities for over 100 CAMO influx intensified video meteors. He found a clear division between meteors with orbits consistent with long-period comets, and those with Jupiter family comet or asteroidal orbits. The former had densities of about 1000 kg m$^{-3}$, while the latter two groups had an average density closer to 4000 kg m$^{-3}$. 

Ablation models which use different fragmentation mechanisms can successfully fit the light curves and deceleration data of meteors, in some cases with very different physical parameter estimates. \citet{campbell2013} used a thermal disruption model and a thermal erosion model to fit ten meteor light curves and decelerations, and then predicted the high-resolution behaviour of the fragments. The two models made very different predictions about the amount and brightness of the wake of the meteors, and neither model was particularly successful at matching the actual wake seen in the CAMO tracking system. High resolution observations of meteor wake provide a strong constraint on ablation models, since many fewer fragmentation mechanisms can match both the wide field data and the wake simultaneously.

\section{Discussion}

Large asteroidal meteors seem to have strengths which exceed the strength of rubble piles, but are less than the strengths of recovered meteorites. This implies that these objects are fractured stones, or that cohesive forces in rubble piles are larger than predicted by current theory. Most cometary meteoroids are much more fragile than asteroidal meteoroids, but some meteoroids and some meteoroid component grains have much larger strengths. These strong cometary objects are of great interest, and determining their origin and composition would be of great interest.

There is ample evidence that small meteoroids are very weakly bound clusters of grains which separate before or during ablation. Even meteoroids from the same parent body show differences in fragmentation behaviour, spectral evolution and begin heights, implying that there is considerable inhomogeneity in the parent bodies. 

There is still a great deal to be learned about the fundamental structure of meteoroids, and high-resolution systems which can put strong constraints on their ablation behaviour will play an important role. Of greatest interest is the size distribution of grains in fragile meteors, and the grain and bulk densities of meteoroids.  Structure may also reveal differences in meteoroids of different ages, since exposure to radiation may change the coherence of small particles. Similarly, close approaches to the Sun are of great interest, since intense heating may alter the mineralogy. Finally, differences among and within parent bodies may point to different formation processes or inhomogeneities in asteroids or comets.

\section*{Acknowledgements}

Thanks to the two anonymous reviewers for their helpful comments which improved the paper.





\end{document}